\documentclass[12pt,preprint]{aastex}

\def\ni{\noindent}

\def\cm{{\rm\,cm}}

\def\gm{{\rm\,g}}

\def\AU{{\rm\, AU}}
\def\mum{\,\mu{\rm m}}

\def\yr{{\rm\,yr}}

\begin{document}

\shortauthors{Chiang and Murray-Clay}
\shorttitle{Circumbinary Ring}

\title{The Circumbinary Ring of KH 15D}

\author{Eugene I.~Chiang \& Ruth A.~Murray-Clay}

\affil{Center for Integrative Planetary Sciences\\
Astronomy Department\\
University of California at Berkeley\\
Berkeley, CA~94720, USA}

\email{echiang@astron.berkeley.edu, rmurray@astron.berkeley.edu}

\begin{abstract}
The light curves of the pre-main-sequence star KH 15D
from the years 1913--2003 can be understood if the star is a member
of an eccentric binary that is encircled
by a vertically thin, inclined ring of dusty gas. Eclipses occur whenever
the reflex motion of a star carries it behind
the circumbinary ring; the eclipses occur with period equal
to the binary orbital period of 48.4 days. Features of the light
curve---including the amplitude of central reversals
during mid-eclipse, the phase of eclipse with respect to the binary
orbit phase, the level of brightness out-of-eclipse, the depth of eclipse,
and the eclipse duty cycle---are
all modulated on the timescale of nodal regression of
the obscuring ring, in accord with the historical data.
The ring has a mean radius near 3 AU and a radial width
that is likely less than this value. While the inner boundary
could be shepherded by the central binary, the outer boundary
may require an exterior planet to confine it against viscous spreading.
The ring must be vertically warped
to maintain a non-zero inclination. Thermal pressure gradients
and/or ring self-gravity can readily enforce rigid precession.
In coming years, as the node
of the ring regresses out of our line-of-sight towards the binary,
the light curve from the system should cycle approximately
back through its previous behavior. Near-term observations should
seek to detect a mid-infrared excess from this system; we estimate
the flux densities from the ring to be $\sim$3 mJy at wavelengths
of 10--100$\mum$.
\end{abstract}


\section{INTRODUCTION}
\label{intro}

The light curve of the pre-main-sequence star KH 15D,
first brought to prominence by Kearns \& Herbst (1998), promises
to yield unique insights into the evolution of young stars
and their immediate environments. Every 48.4 days, the star undergoes
an eclipse, about which we know the following:

\begin{enumerate}
\item{Between 1995 and 2003,
the eclipse duty cycle (fraction of time spent in eclipse)
has grown from 30\% to 45\%
(Hamilton et al.~2001; Winn et al.~2003).}

\item{During these years, ingress and egress each occupy
2--3 days out of every cycle (Herbst et al.~2002).}

\item{The in-eclipse light curve during these years exhibits
a central reversal in brightness (Hamilton et al.~2001; Herbst et al.~2002).
The amplitude of the reversal
has lessened with time. In 1995, when the central
reversal was first observed, the peak brightness of the reversal
exceeded the out-of-eclipse brightness.}

\item{Light from the star in mid-eclipse is linearly polarized
by a few percent across optical wavelengths, suggesting that
a substantial fraction of the light in-eclipse is scattered
off dust grains whose sizes exceed a few microns (Agol et al.~2004)}.

\item{From 1967--1982, the system underwent eclipses with the
same 48.4 day period as in recent years, with a
duty cycle of $\sim$40\%. In contrast to its out-of-eclipse
state today, its out-of-eclipse state then
was brighter by $\sim$0.9 magnitudes. Moreover, the eclipse
was less deep---only $\sim$0.7 mag deep then as compared
to today's maximum depth of 3.5 mag. Its phase then was also shifted
by $\sim$0.4 relative to today (Johnson \& Winn~2004).}

\item{From 1913--1951, no eclipse was observed (Winn et al.~2003).}
\end{enumerate}

We present here a physically grounded picture in which all
of these observations can be understood. Its most basic
elements are described in \S\ref{model}, where we demonstrate that
the various timescales exhibited by the light curve can
be explained by an inclined, vertically thin, nodally
regressing ring of dusty gas that surrounds a stellar
binary of which KH 15D is one member. 
For the ring plane to
maintain a non-zero inclination with respect to the binary plane,
it must be vertically warped (not merely flared;
i.e., the mean inclinations of ring streamlines
with respect to the binary plane must vary across the ring)
so that thermal pressure gradients or
ring self-gravity can offset the differential nodal precession
induced by the central binary. The most natural geometry
for the ring is that it be radially narrow; by analogy
with narrow planetary rings
that are accompanied by confining shepherds, we suggest
that a planetary companion orbits exterior to
the circumbinary ring of KH 15D.
Perhaps the chief attraction of the model lies
in its ability to make predictions;
these predictions
are also described in \S\ref{model}. A summary of our model,
a discussion of the significance KH 15D carries in our
overall understanding of the evolution of circumstellar,
presumably protoplanetary disks, and a listing of directions
for future research, are contained in \S\ref{discussion}.

\section{MODEL}
\label{model}

\subsection{Basic Picture and Model Light Curves}

Motivated by (1) significant radial velocity variations
of KH 15D as measured by Johnson et al.~(2004, submitted),
(2) the observation of a central reversal
in 1995 for which the peak brightness exceeded the out-of-eclipse
brightness, and (3) the systematically greater brightness of
the system in 1967--1982 as compared to recent years, we consider
the pre-main-sequence K star KH 15D to possess an orbital companion.
The data described by Johnson \& Winn (2003) are consistent
with a companion (hereafter, K$'$) whose luminosity is $\sim$20\% greater
than that of KH 15D (hereafter, K). All quantities superscripted with
a prime refer to the orbital companion of KH 15D.
The mass of K$'$ should be nearly
the same as that of K, and we assign each a mass of
$m_b = m_b' = 0.5 M_{\odot}$ consistent with the system's T Tauri-like
spectrum.
We identify the eclipse period of 48.4 days with the orbital period of
the binary; for our chosen parameters, the semi-major
axis of each orbit referred to the center-of-mass is $a_b = a_b' = 0.13$ AU.
We assign an orbital eccentricity of $e_b = 0.5$ based on a preliminary
analysis of data taken by Johnson et al.~(2004).
The precise value is not important;
the only requirement
is that the orbital eccentricity be of order unity.

The eclipses are caused by an annulus of dust-laden gas that encircles
both stars, beginning at a distance $a_i > a_b$ as measured from
the binary center-of-mass, and ending at an outer radius
$a_f = a_i + \Delta a$.
The symmetry plane
of the ring is inclined with respect to the binary plane
by $\bar{I} > 0$. We defer to \S\ref{rigid}
the issue of how such a ring maintains
a non-zero $\bar{I}$
against differential nodal precession.
The ring will nodally regress at an angular speed of

\begin{equation}
\dot{\Omega} \sim -\bar{n} \left( \frac{a_b}{\bar{a}} \right)^2 \sim -0\fdg 13
\yr^{-1} \left( \frac{\bar{a}}{3 \AU} \right)^{-7/2} \, ,
\label{regress}
\end{equation}

\ni where $\bar{a}$ is the mean radius of the ring and $\bar{n}$
is the mean motion evaluated at that radius. Equation (\ref{regress}),
derived from standard celestial mechanics perturbation theory,
is only accurate to order-of-magnitude, since it relies on an expansion
that is only valid to first order in $m_b'/m_b$. Nonetheless, it is
sufficiently accurate to establish the reasonableness of our picture
in the context of the observations; corrections will not alter
our conclusions qualitatively.

\begin{figure}
\epsscale{0.6}
\plotone{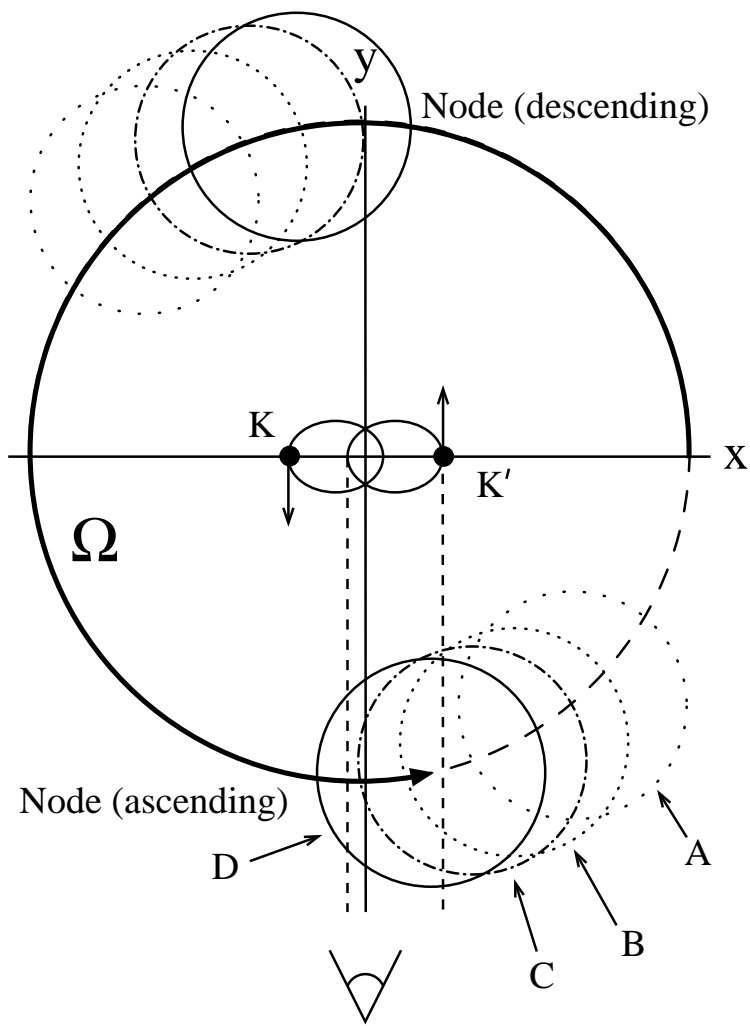}
\caption{Schematic of the KH 15D system (not to scale).
The star KH 15D is denoted K, while its orbital companion is denoted
K$'$. The two stars are of nearly the same mass and
occupy highly eccentric orbits. Surrounding the binary
is a dusty ring, whose ascending and descending nodes
(``footprints'') on the binary plane are indicated by smaller circles.
The sizes of the circles represent the amount of obscuring material
viewed along our line-of-sight in the binary plane; the sizes
are set by the ring's inner and outer radii,
the degree of vertical flaring, and
the inclination profile (mean inclination plus warp).
The quadrupole field of the central
binary causes the ascending node of the ring to regress (travel
counter to the direction of orbital mean motion) from position
A to position D. The longitude of ascending node is measured
counter-clockwise from the x-axis and is denoted by $\Omega$.
The observer views the binary orbit edge-on from below.
More detailed schematics of the ring-binary geometry corresponding
to phases A--D can be found in Figure \ref{f2}.}
\label{f1}
\end{figure}

As illustrated in Figure \ref{f1}, eclipses occur
whenever the ascending or descending node of the ring
regresses into our line-of-sight towards the stellar binary.
The observer is assumed to view the binary orbit edge-on, or nearly so.
The orbital motion of a given star about the center-of-mass
causes the star to be occulted by varying columns of ring material.
Eclipses occur with a period equal to the binary
orbital period. The shape of the eclipse---e.g., the presence
or absence of central reversals, or the level of brightness
out-of-eclipse---is modulated over the longer timescale of nodal
regression.

\placefigure{fig1}

\begin{figure}
\epsscale{0.9}
\plotone{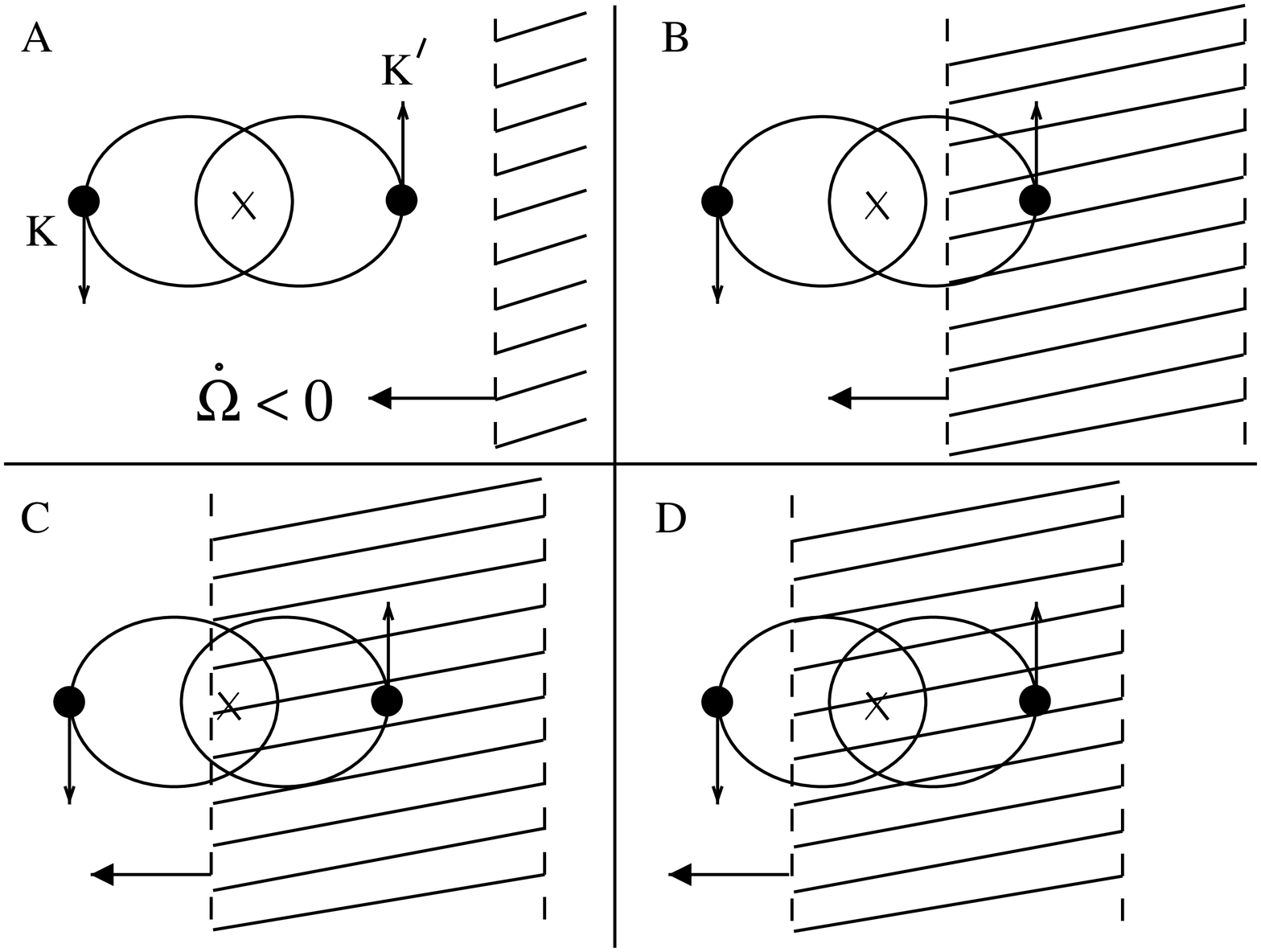}
\caption{Schematic of relative ring-binary geometries.
The inset letters in each panel
correspond to the same geometries depicted
in Figure \ref{f1}; the observer views the binary from the bottom
of the page. In panel A, no eclipses occur,
consistent with the historical data prior to 1951.
In panel B, the ring's node has regressed to block the apoapsis
of star K$'$; eclipses like those seen from 1967--1982 should
be seen. In panel C, eclipses like those seen from 1995--2000
should be evident, with central reversals of brightness
during mid-eclipse when star K$'$ emerges at its periapsis.
In panel D, the node covers the entire orbit of K$'$
and an increasingly large fraction
of the orbit of K; central reversals should be much weaker,
and the eclipse duty cycle greater, than in C. While
we have drawn the occulted region as a hatched region of definite area,
the intervening optical depth along different rays is a smooth function,
both in reality and in our computations of the light curve.
}
\label{f2}
\end{figure}

Figure \ref{f2} depicts schematically and in more detail some of the
ring-binary geometries that are possible. Each panel in Figure \ref{f2}
is marked with a letter corresponding to a particular longitude of
ascending node, $\Omega$, of the ring on the binary plane; the letters
in Figures \ref{f1} and \ref{f2} correspond to the same geometries.
For example, in panel B of Figure \ref{f2}, when the edge of the ring
occults the apoapsis of the orbit of K$'$, the observer should see
eclipses like those witnessed in 1967--1982; star K is always seen,
while star K$'$ periodically disappears behind the ring; out-of-eclipse,
light from two stars is seen, while in mid-eclipse, only one star is seen.
In panel C, as the ring's edge regresses further inwards
to just cover the binary center-of-mass, only the periapsis of K$'$
and the apoapsis of K remain unobscured; during the eclipse of K,
the brighter companion K$'$ emerges briefly, producing a central
reversal like that seen in 1995. Other panels in Figure \ref{f2}
correspond to other longitudes of the node, and can be
identified with other observed behaviors of the light curve.
In particular, the regression of the ring's edge onto and past
the periapsis of K$'$ (panel D) results in a
concomitant weakening of the central reversal
and a lengthening of the duration of the eclipse of K---trends
seen today.

\placefigure{fig2}

A key parameter of the ring is its vertical scale height, which must
be small enough to yield short ingress and egress times, but large
enough to cover substantial fractions of the binary orbit.
We model the ring with a Gaussian atmosphere perpendicular to its
midplane at a given radius, as befits material with constant
vertical velocity dispersion. We describe the absorption coefficient (units
of inverse length) by

\begin{equation}
\alpha = \alpha_i \left( \frac{a}{a_i} \right)^{-\Gamma} \exp \{- [\theta(a) /
\theta_0]^2 \} \, ,
\label{gaussian}
\end{equation}

\ni where $\theta(a)$ is the latitudinal angle
measured from the {\it local} ring plane at disk radius $a$,
and $\theta_0$, $\alpha_i$ and $\Gamma$ are constants.
We prescribe a power law for
the inclination of the local ring plane with respect
to the binary plane:

\begin{equation}
I(a) = I_i \left( \frac{a}{a_i} \right)^{\beta} \, ,
\end{equation}

\ni where $I_i$ and $\beta$ are constants.
The function $I(a)$ specifies the vertical {\it warp} across the ring,
as distinct from the finite thickness described by the Gaussian
in equation (\ref{gaussian}).\footnote{A warp is not the same
as a flare. The latter term refers to an increasing $\theta_0$
with disk radius $a$.}
A warp must be present to maintain rigid nodal precession; see section
\S\ref{rigid}.
Our standard choices for model parameters
are contained in Table \ref{t1}. These are chosen to reproduce
approximately the observed light curves, as we describe below.

Computation of the light curve in the absence of scattering is a
straightforward exercise in numerical integration. For simplicity,
and because our purpose here is to introduce new ideas,
we model the stars as point sources.
Light from each star
is extinguished as $\exp (-\int \alpha \, dl)$, where the integral
is performed along a ray from the star to the observer.
The orientation of the observer is fixed such that the binary orbit
is viewed edge-on and the velocity of K at its apoapsis
is directed towards the observer; see Figure \ref{f1}.

\begin{figure}
\epsscale{0.9}
\plotone{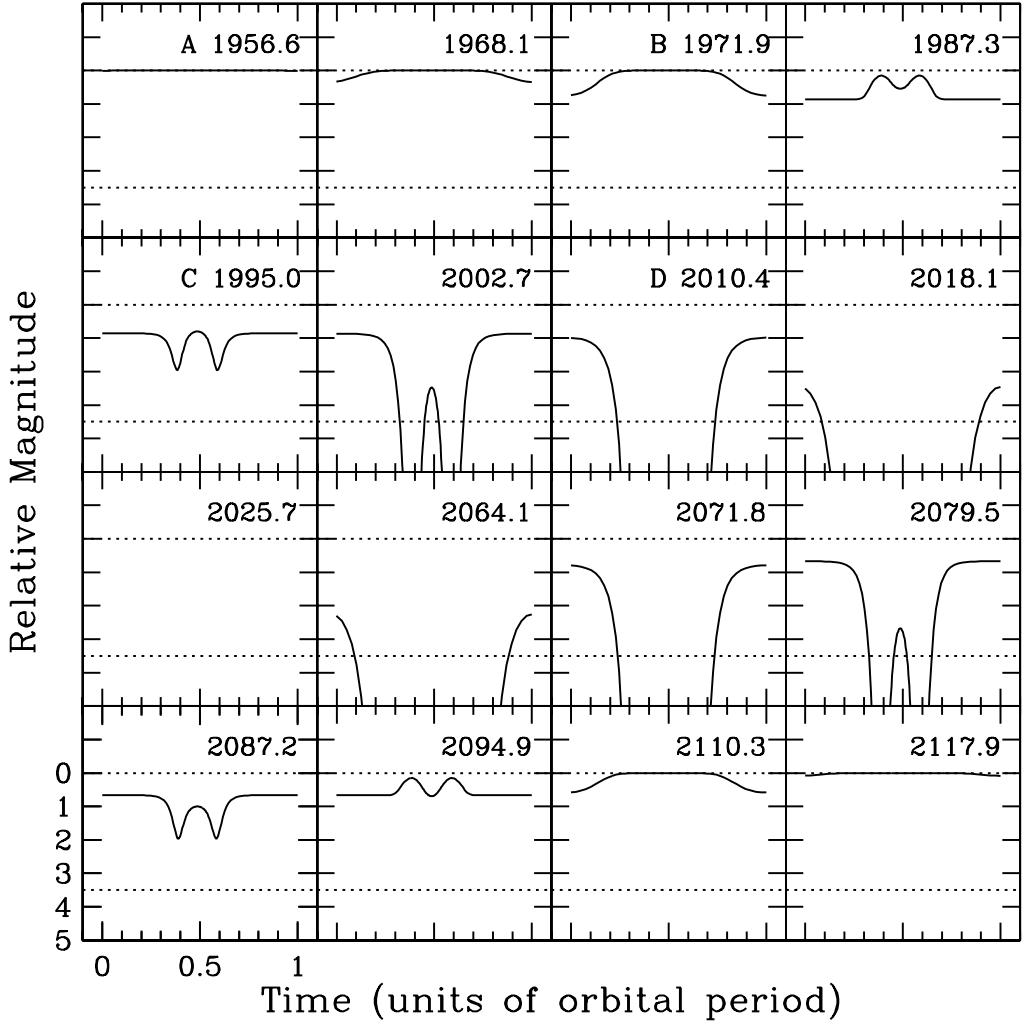}
\caption{Computed light curves for the KH 15D system,
using parameters listed under model 1 of Table \ref{t1}.
Inset letters A--D refer to the same geometries
illustrated in Figures \ref{f1} and \ref{f2}.
The inset number refers to the year in which
behavior similar to that computed was
actually observed; after we assign the year 1995.0
to panel C in which the strongest central flash
is evident, years for all other panels
are computed according to the nodal
regression rate [equation (\ref{regress})]
and the value of  $\Omega$ appropriate to a given panel.
The upper dashed line in each panel indicates the
maximum system brightness; the lower dashed line
indicates a possible minimum system brightness
established by light scattered off circumbinary material.
Predictions for future years are shown; roughly speaking, the system
should cycle through its previous behavior in reverse order.}
\label{f3}
\end{figure}

Figure \ref{f3} displays a sequence of light curves
for parameters corresponding to model 1 in Table \ref{t1}.
Each panel corresponds to a particular choice of $\Omega$;
this angle decreases from $\Omega = 282^{\circ}$
to $\Omega = 259^{\circ}$ from top left to bottom right.
Four panels are labelled with letters according to the same
scheme used in previous figures. We assign the year 1995.0
to panel C, in which the peak brightness of the central reversal
exceeds the out-of-eclipse brightness. Dates for all other
panels are computed according to equation (\ref{regress}).

We are encouraged by the agreement between Figure \ref{f3} and
the observations as summarized in \S\ref{intro}.
A primary shortcoming of our computations
is the neglect of starlight scattered off the circumbinary ring
and potential other material, such as a remnant envelope.
It is our hope that scattered light sets a lower limit to the observed
flux, $\sim$3.5 magnitudes below the magnitude of K, as suggested
by Agol et al.~(2003). We indicate such a floor by the lower dashed line
in every panel of Figure \ref{f3}.

\placefigure{fig3}

Predictions for the light curve for a given set of model parameters
are readily computed. In the near future, 
the system should simply cycle backwards through its previously
observed behavior, as the trailing edge of the node sweeps across
the binary orbit.\footnote{If the observer views the binary orbit
exactly edge-on as shown in Figure \ref{f1},
it is unclear from the light curve whether
the ascending or descending node is occulting the binary. The
degeneracy can be broken for other viewing geometries.}
The evolution will not, however, be
exactly mirror-symmetric;
for example, in later years, we do not expect the central reversal
to exceed the out-of-eclipse brightness, since K is
less luminous than K$'$; compare, in Figure \ref{f3},
the bottom left panel to panel C.
Changes in viewing orientation will also alter details.

We defer to modellers of more sophistication than ourselves the task
of rigorously fitting the observed light curves to achieve predictions
of greater robustness. None of the parameters in Table \ref{t1}
should be considered unique; for example, we generated
light curves similar to those shown in Figure \ref{f3} by
reducing $I_i$ and $\theta_0$ simultaneously by factors of 10.

We turn now to two related
theoretical issues: the ability of the ring to maintain
a non-zero inclination,
and the determination of its inner and outer radii.

\subsection{Maintaining a Non-Zero Inclination}
\label{rigid}
In the absence of thermal pressure and ring self-gravity, a circumbinary
disk of particles, initially inclined with respect to the binary
plane and characterized by the same longitude of ascending node over all
radii,
would precess differentially; radial variations in
the nodal precession rate due to the quadrupole field of the central
binary would reduce the mean inclination of the disk to zero.
Narrow planetary rings that are observed to maintain non-zero
inclinations with respect to the equator planes of their host planets
avoid this fate by virtue of their self-gravity, with some modification by
pressure gradients
(Chiang \& Culter 2004; Chiang \& Goldreich 2000; Borderies, Goldreich \&
Tremaine 1983).
Gaseous disks can also maintain rigid precession by the action
of thermal pressure (Papaloizou \& Pringle 1983; Papaloizou \& Lin 1995;
Larwood \& Papaloizou 1997; Lubow \& Ogilvie 2000;
Lubow \& Ogilvie 2001) and/or self-gravity;
some indirect evidence for rigid precession in gaseous, circum{\it primary}
disks is available from X-ray binaries (Larwood 1998) and pre-main-sequence
binaries (Terquem et al.~1999).

Whether the ring is composed
of particles or gas, it must be vertically
warped so that either pressure or self-gravity
exerts forces normal to the local ring plane. That is,
given a ring in which each streamline has the same longitude
of ascending node as every other streamline, there must
exist a gradient in inclination across the ring.

Unlike the case of planetary rings, thermal pressure alone is sufficient to
offset differential precession in gaseous, circumstellar disks
because the gas sound speed, $c_s$, is typically a healthy fraction of the
Kepler orbital speed, $\bar{n}\bar{a}$.
To maintain rigid precession across a circumbinary ring
of radial width $\Delta a < \bar{a}$ by gas pressure alone,
the ring must exhibit a fractional variation in inclination of order

\begin{equation}
\frac{\Delta I}{\bar{I}} \sim -0.1 \left( \frac{\bar{n}\bar{a}/c_s}{20}
\right)^2 \left( \frac{a_b/\bar{a}}{0.05} \right)^2 \left( \frac{\Delta
a/\bar{a}}{0.5} \right)^3 \, ,
\label{press}
\end{equation}

\ni where the numerical evaluation is appropriate for parameters listed
under model 1 of Table \ref{t1}. The narrower the ring, the less
severe is the requisite warp. The inclination gradient is
negative; from its inner edge to its outer edge, the ring
tends to bend back down towards the binary plane. This order-of-magnitude
expression can be derived by setting the differential precession
rate between two infinitesimally narrow wires due to the quadrupole
field of the central binary equal to the differential precession
rate due to the repulsive pressure between the wires; the spirit
of the calculation is the same as can be found in Goldreich \& Tremaine
[1979a, see their equations (12)--(14); or 1979b].
Equation (\ref{press}) must be viewed with caution; it neglects pressure
forces
at ring boundaries that could be important, especially
if the pressure changes over lengthscales that are
shorter than $\Delta a$ (such as the vertical scale height
$h$; see Chiang \& Goldreich 2000).
It also assumes zero radial gas flow between disk annuli; Lubow
\& Ogilvie (2004, personal communication) have shown that by accounting
for the near-resonant forcing of radial velocities
by gas pressure (Papaloizou \& Pringle 1983; Papaloizou \& Lin 1995),
the magnitude of the warp might be reduced from that
predicted by equation (\ref{press}). Definitive models of the ring
will require these issues to be addressed; for now,
we feel these concerns will not alter our qualitative
conclusions.

If rigid precession is maintained by self-gravity alone,
the magnitude of the warp is a function of the ring mass, $m_r$:

\begin{equation}
\frac{\Delta I}{\bar{I}} \sim 0.1 \left( \frac{m'_b/m_r}{500} \right) \left(
\frac{a_b/\bar{a}}{0.05} \right)^2 \left( \frac{\Delta a/\bar{a}}{0.5}
\right)^3 \, ,
\label{sg}
\end{equation}

\ni where the numerical evaluation is appropriate for the parameters
listed under model 2 of Table \ref{t1}. 
Here the inclination gradient
is positive. The same steep dependence on ring width is evident.
Our chosen ring mass, equal
to the mass of Jupiter, is comparable to that expected from the
minimum-mass solar nebula and still yields
%Our choice of normalization
%for $m_r/m'_b = 1/500$ still yields
a ring that is gravitationally stable in the
Toomre $Q$ sense; the ring mass divided by the central mass must approach
$c_s / \bar{n}\bar{a} \sim 1/20$ before gravitational instability in the
gas becomes relevant.

If $m_r/m'_b \ll 1/500$, then equation (\ref{press}) better describes
the warp, to the extent that it is difficult to imagine values
for $c_s$ much different from that assumed. If $m_r / m'_b \gg 1/500$,
then equation (\ref{sg}) better describes the warp.

\begin{figure}
\epsscale{0.9}
\plotone{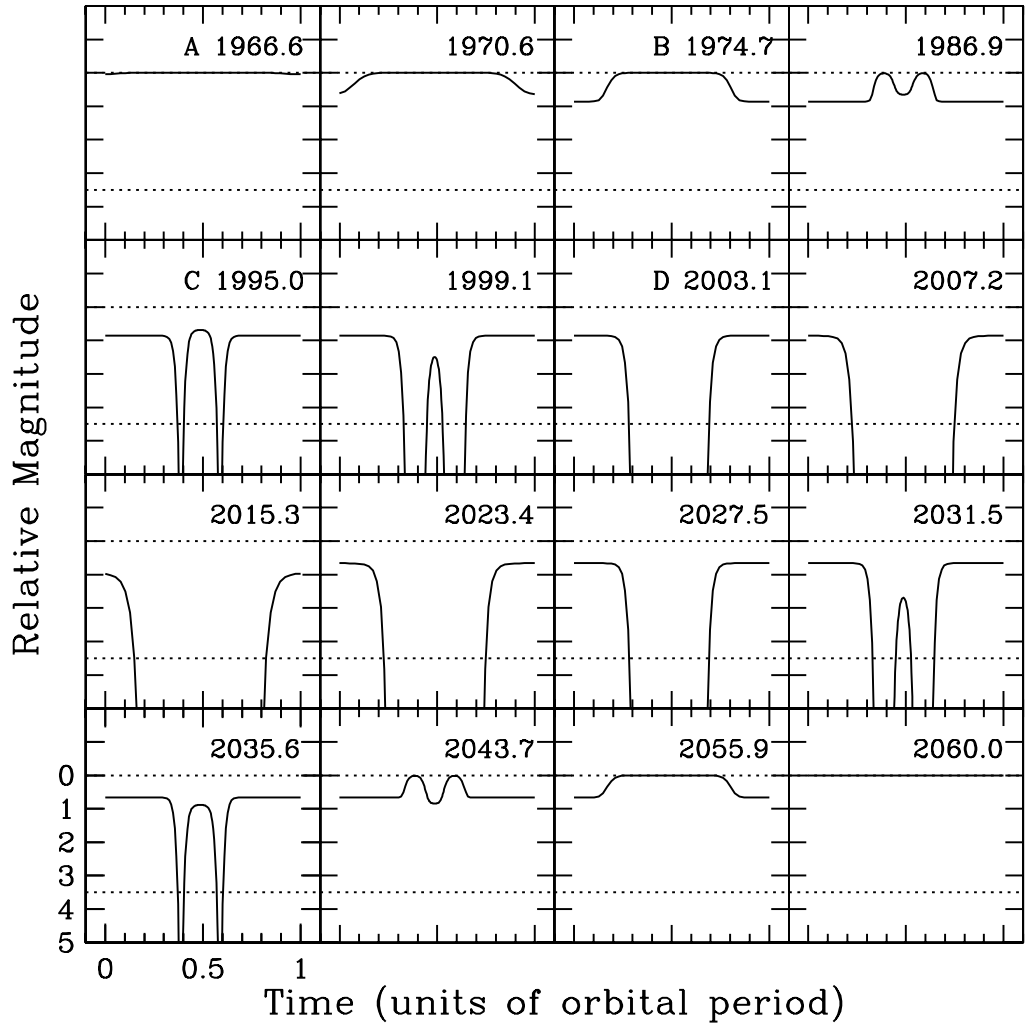}
\caption{Same as Figure \ref{f3}, but for model 2,
for which the inclination gradient across
the ring is positive. The evolution of light curves is qualitatively the
same as that for model 1, for which the inclination gradient
is negative.}
\label{f4}
\end{figure}

Models 1 and 2 are intended to represent disks for which
thermal pressure and self-gravity, respectively, are wholly
responsible for maintaining a non-zero $\bar{I}$.
In model 1, the inclination of a streamline decreases from $I(a_i) =
20^{\circ}$ to $I(a_f) = 10^{\circ}$; in model 2, $I(a_i) =
10^{\circ}$ and $I(a_f) = 20^{\circ}$. The inclination
profiles so prescribed accord with equations (\ref{press}) and (\ref{sg}).
Light curves from model 1 are displayed in Figure \ref{f3},
while those from model 2 are showcased in Figure \ref{f4}. We consider
the agreement with the observations to be comparable between the two
cases. Note that for model 2,
the binary orbit is less severely occulted, in contrast to model 1.
Moreover, the ingress and egress profiles are sharper in model 2.

\placefigure{fig4}

\subsection{Radial Boundaries}
\label{edges}

\subsubsection{The Outer Boundary}

If this dusty disk extends too far outwards in the radial direction
($\Delta a \gg a_i$), at least two problems arise: (1) how to
maintain rigid precession across an extended disk, and (2) how
to prevent a disk of large vertical thickness from occulting the central
binary completely at all times. We discuss each of these in turn.

``Rigid tilt modes'' can be
described for disks that extend for more than one decade in radius (Lubow \&
Ogilvie 2004, personal communication). Gas pressure can enforce such a mode
over a length travelled by a sound wave in a differential
precession time ($\sim$20 AU for a sound speed
of $\sim$1 km s$^{-1}$ and a differential precession time
of $\sim$10$^2$ yr evaluated at 3 AU). Self-gravity could extend this length.
It is unclear why such a mode would dominate a large,
extended disk; on the other hand, while it is similarly
not understood why narrow planetary rings exhibit rigid tilts, at least clear
observational evidence exists that they do (French et al.~1991).

The second problem could be avoided if the ring is sufficiently
thin and sufficiently warped that material is bent out of
the line-of-sight. Dust need not be well-mixed with gas;
the tidal force of stellar gravity can pull grains towards
the local disk midplane. Evidence for dust settling
is abundant in classical
T Tauri and Herbig Ae systems [see, e.g., Chiang et al.~(2001)],
but the required degree of settling becomes more severe with increasing
disk radius.\footnote{The opening angle of $\theta_0 \sim 0.3$ deg
that we employ in models 1 and 2 for the dust disk
is an order of magnitude lower than the typically assumed
opening angle of the gas disk at this radius. This low value,
presumably the result of dust settling, still sits an order of magnitude
above the minimum opening angle set by Richardson turbulence.}

While the above problems can be surmounted to
varying degrees, we still feel
drawn to a radially narrow ring of dust as the most
natural solution. Our assumptions regarding the central stellar
binary, together with the requirement that the ring's nodal precession rate
accord with the observed history of light curves,
set the mean radius of the ring to be $\bar{a} \approx 3$ AU.
Having experimented with a variety of model parameters,
we feel no more precise statement can be made other than
that $\Delta a \lesssim \bar{a}$.

What could be responsible for such a ring's confinement at its
outer radius? Perhaps the
most obvious proposal to make is to draw an analogy with narrow planetary
rings that are radially confined by shepherd moons
(see, e.g., Goldreich \& Tremaine 1982). These rings successfully
maintain non-zero inclinations with respect to the equator
planes of their central planets. Unfortunately, mass estimates
for shepherding planets in the case of KH 15D cannot be made without
knowing the viscosity of disk material; the viscosity could
easily be so tiny that planets less massive than Jupiter could confine
the ring against viscous spreading.

An alternative to confinement by a planetary
companion could be provided by radial migration of dust due to aerodynamic
drag
(for the background physics, see, e.g., Youdin \& Shu 2002;
Youdin \& Chiang 2004; and references therein). In this picture,
dust accretes relative to gas
towards the inner edge of the disk, leaving behind dust-depleted gas.
The accretion rate depends on grain size, however, and it seems
implausible to expect a grain size distribution that is so narrowly
peaked in a given size range that the disk
nearly completely empties itself of radial
optical depth outside $a \sim a_i$.

\subsubsection{The Inner Boundary}

The central stellar binary itself could
shepherd the inner edge of the ring. Artymowicz \& Lubow (1994)
compute $a_i/a_b \approx 5$ for the case of binaries
having $e_b \approx 0.5$ and $m'_b/m_b \approx 0.3$, and
moderately viscous disks. This would place $a_i \approx 0.6$ AU.
The true value for the case of KH 15D would be greater than this since the
binary mass ratio is closer to unity.

An AU-sized inner hole can naturally explain
the observed absence of near-infrared excess for KH 15D.
Such excesses arise from reprocessing
of starlight by and/or active accretion of disk material
at stellocentric distances inside a few AU.
By carving out an AU-sized inner hole in
the disk surrounding the T Tauri star TW Hydra,
Calvet et al.~(2002) succeed in reproducing
the observed deficit of near-infrared emission
in its spectral energy distribution.
For TW Hydra, disk material at stellocentric distances greater
than an AU produces excess emission at wavelengths longward
of 10 microns (see also Weinberger et al.~2002).
We hold the same expectation for KH 15D,
as described quantitatively in the next section.

\section{SUMMARY AND DISCUSSION}
\label{discussion}

We have proposed that the light curves of the pre-main-sequence
star KH 15D can be understood if that star harbors a companion
of slightly greater luminosity and nearly identical mass, on an
orbit having eccentricity of order unity. Today, this companion is
occulted by a circumbinary ring of dusty gas. To cover one star
and not the other, the ring
is necessarily inclined with respect to the binary plane;
to maintain uniform nodal precession and a mean inclination,
it is also necessarily warped.\footnote{The ring is likely
to be eccentric as well, with a mean eccentricity induced
by secular forcing from the binary.}
Thermal pressure gradients or self-gravity readily
furnish the forces necessary to maintain rigid precession;
the inclination gradient across the ring is negative
or positive depending on whether pressure or gravity dominates.
While a variety of ring geometries appear to give fits
of comparable quality to the light curves, the
following dimensions to the ring seem difficult
to avoid: a mean radius of $\bar{a} \approx 3$ AU,
and a radial width $\Delta a \lesssim \bar{a}$.
Modelling the scattered optical light from the system will yield
constraints on the ring geometry that we have been unable to provide.

\begin{figure}
\epsscale{0.8}
\plotone{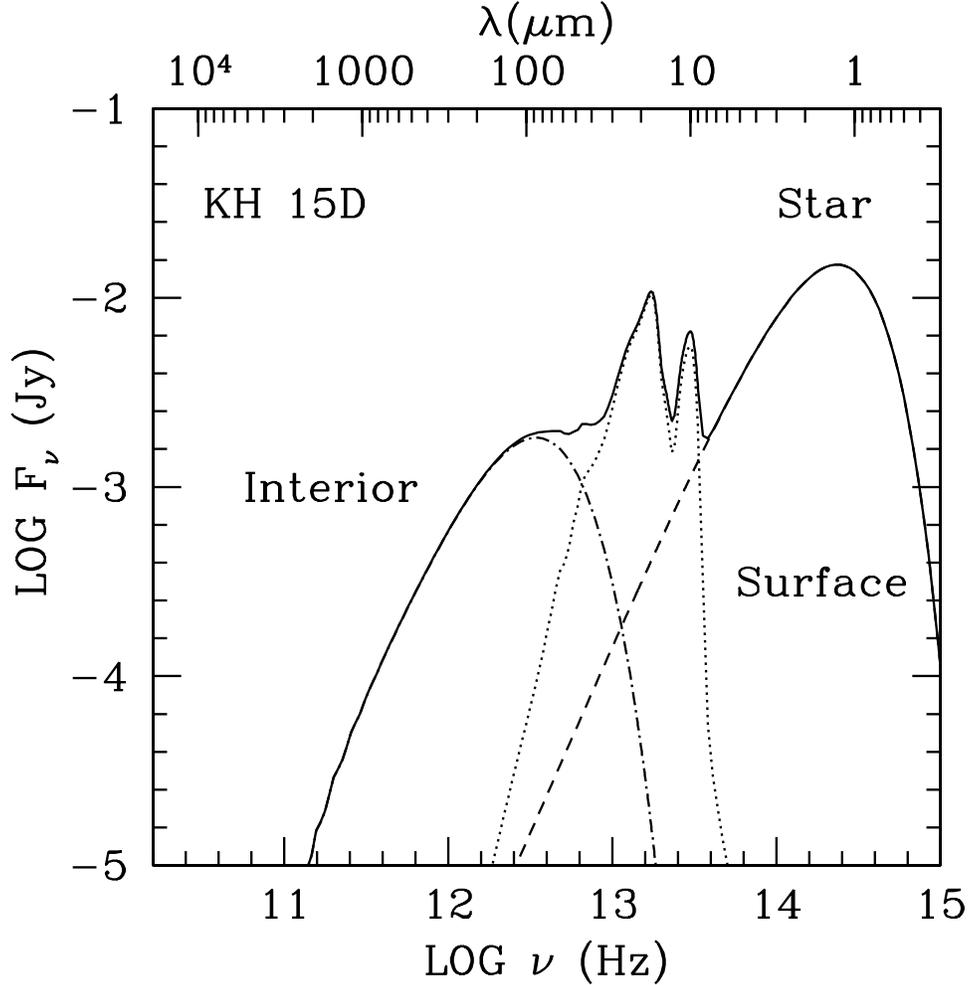}
\caption{Predicted flux density of thermal emission from the KH 15D system,
crudely estimated using the formalism of Chiang et al.~(2001, hereafter C01).
Only emission from one star is shown, though both stars are accounted for
in calculating the reprocessed flux from the ring.
Disk parameters are the same as those of the standard model of C01
except for the inner radius ($a_i = 1$ AU), the outer radius
($a_o = 5$ AU), and the height of the photosphere relative
to the gas scale height ($H/h = 1.4$).
Our choice of $H/h$ simulates a ring warp for which $\Delta I = 1^{\circ}$.
To account for the nearly edge-on viewing geometry,
the flux from the disk interior has been reduced from the
face-on value by a geometric projection factor of 10.
The flux from the optically thin
disk surface has not been adjusted from its face-on value
(see Chiang \& Goldreich 1999).}
\label{f5}
\end{figure}

Looking farther ahead, we should expect an
observable mid-infrared excess from this ring by passive reprocessing
of starlight. To our
knowledge, no mid-infrared observation has been taken
of KH 15D. At wavelengths of 10--$100\mum$, we estimate
flux densities of $F_{\nu} \sim 3$ mJy for the
ring parameters in Table 1, well above stellar photospheric
flux densities. This emission arises largely from the
optically thin surface layers of the disk that are directly
exposed to central starlight (Chiang \& Goldreich 1997).
A crudely estimated spectral energy distribution is displayed
in Figure \ref{f5}.
At wavelengths longer than $\sim$100 microns, the flux
density should decrease, if, as we suspect, the ring
is truncated at an outer radius of a few AUs.
In making this esimate, we assume that only 10\% of full
flux from a face-on, optically thick disk interior is observed
because of a nearly edge-on viewing orientation and that the angle
with which starlight grazes the surface of the warped dust disk
is $\Delta I \sim 1^{\circ}$ (see, e.g., Chiang \& Goldreich 1997, 1999).

\placefigure{fig5}

Observations of the stars in and out of eclipse
at longer wavelengths would enable us to measure
the wavelength dependence of dust opacity and thereby constrain
the size distribution of disk grains. Such occultation observations
are routinely performed for planetary rings from ultraviolet
to radio wavelengths, and have provided a wealth of information
for such systems. Non-thermal radio emission from T Tauri stars
might well be strong enough to provide a background light source.

The period of eclipses is that of the binary orbital period,
while the shape of the light curve is modulated
over the much longer timescale of nodal precession.
We anticipate that in the coming decades, as the ring node
regresses past the binary orbit, the eclipses
will eventually repeat their prior behavior in reverse order:
the eclipses will narrow, the central flashes will amplify,
there will be a shift in eclipse phase by 0.5,
and eventually the eclipses will cease altogether (until the other
node swings by, centuries hence).
The fraction of time during which the circumbinary ring (either
its ascending or descending node)
occults either star is roughly $40^{\circ}/360^{\circ} \sim 11\%$.

The framework we have presented can easily accommodate
fine details of the light curve. For example, slight differences
between ingress and egress time intervals (Herbst et al.~2002)
can be reproduced by orienting the observer at a small angle relative
to quadrature of the binary orbit, thereby taking advantage of the
non-uniform angular motion of a star.

Once the ring's node regresses past the orbit of the companion
of KH 15D over the next few decades, the spectrum of the star can be cleanly
obtained. Measurement of the masses and luminosities
of both components of this pre-main-sequence binary
would provide an important test
of evolutionary tracks on Hertzprung-Russell diagrams. 

The requirement of a radially narrow ring
suggests confinement by tidal torques. 
The central binary could provide the torque required
to prevent viscous spreading of the inner edge
at $a_i \sim 1\AU$.
Shepherding the outer edge a few AUs away would require
a circumbinary planet. If this picture is correct,
then the epoch of planet formation
could not have lasted more than the age of this
T Tauri star, a few million years (Sung, Bessel, \& Lee 1997).

\acknowledgements
Many thanks to John Johnson and Geoff Marcy for generously
sharing their preliminary radial velocity results with us.
We are grateful to Steve Lubow and Gordon Ogilvie for
broadening discussions, Drake Deming for motivating us
to calculate the thermal emission from the ring,
Eric Ford for preliminary calculations of scattered light,
and an anonymous referee for improving
the presentation of this paper.
E.I.C. acknowledges support by
National Science Foundation Planetary Astronomy Grant 
AST 02-05892 and Hubble Space Telescope Theory Grant HST-AR-09514.01-A.

\newpage
\begin{deluxetable}{ccc}
\tablewidth{0pc}
\tablecaption{Model Ring Parameters\label{t1}\tablenotemark{a}}
\tablehead{
\colhead{Parameter}  &
\colhead{Model 1\tablenotemark{b}} & \colhead{Model 2\tablenotemark{c}}
}
\startdata

$\bar{a}$ (AU) & 3 & 2.5 \\

$\Delta a$ (AU) & 1.50 & 1.00  \\

$\alpha_i$ (AU$^{-1}$)\tablenotemark{d} & 50000 & 50000 \\

$\Gamma$ & 4.0 & 4.0 \\

$\theta_0$ (deg) & $0.30$ & $0.35$ \\

$I_i$ (deg) & 20 & 10 \\

$\beta$ & -2.0 & 1.7 \\
\enddata
\tablenotetext{a}{None of the parameters
in this table should be regarded as uniquely
fitting the circumbinary ring of KH 15D;
our only secure conclusions are that $\bar{a} \approx 3$ AU
and $\Delta a \lesssim 3$ AU; see text.}
\tablenotetext{b}{Model 1 represents a disk whose non-zero
inclination is maintained by thermal pressure; the warp index
$\beta$ is negative
so that the ring tilts increasingly towards the binary plane
from small to large radius.}
\tablenotetext{c}{Model 2 represents a disk whose non-zero
inclination is maintained by self-gravity; the warp index
$\beta$ is positive
so that the ring tilts increasingly away from the binary plane
from small to large radius.}
\tablenotetext{d}{A value of $\alpha_i = 50000/$AU
is consistent with an opacity of $1 \cm^2/ (\gm \,{\rm of~gas})$
and a midplane gas density of $3 \times 10^{-9} \gm/\cm^3$.}
\end{deluxetable}

\end{document}